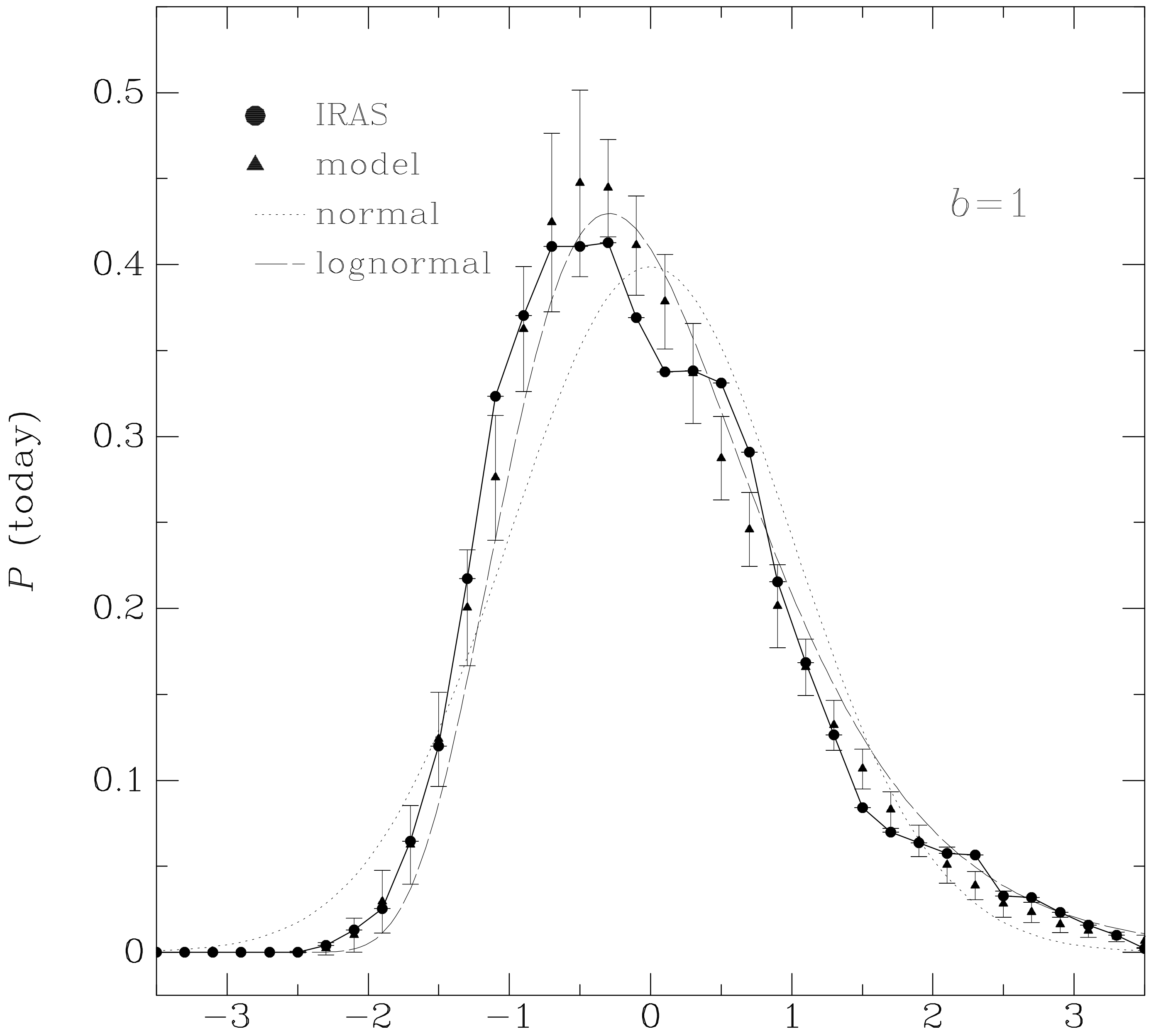

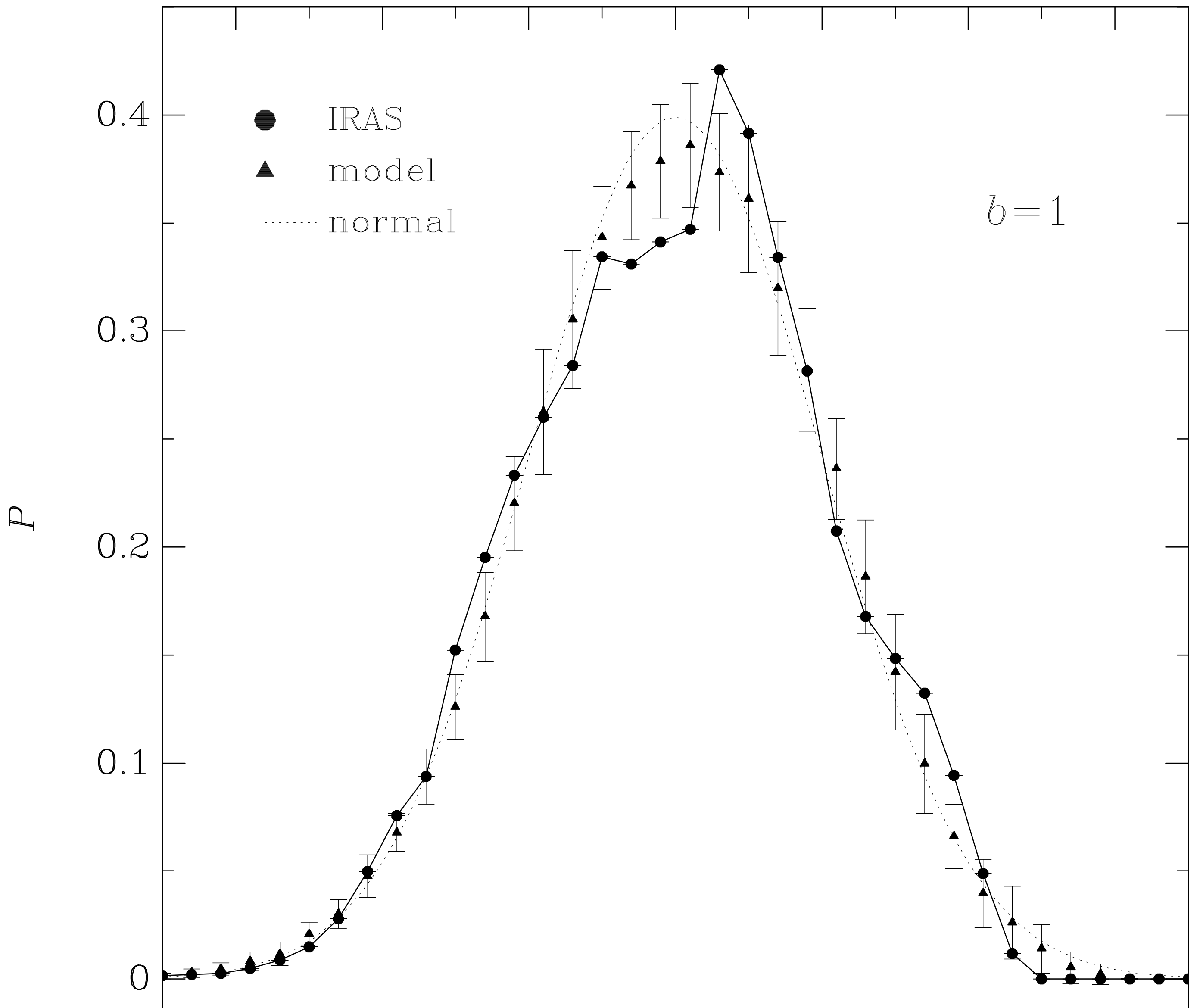

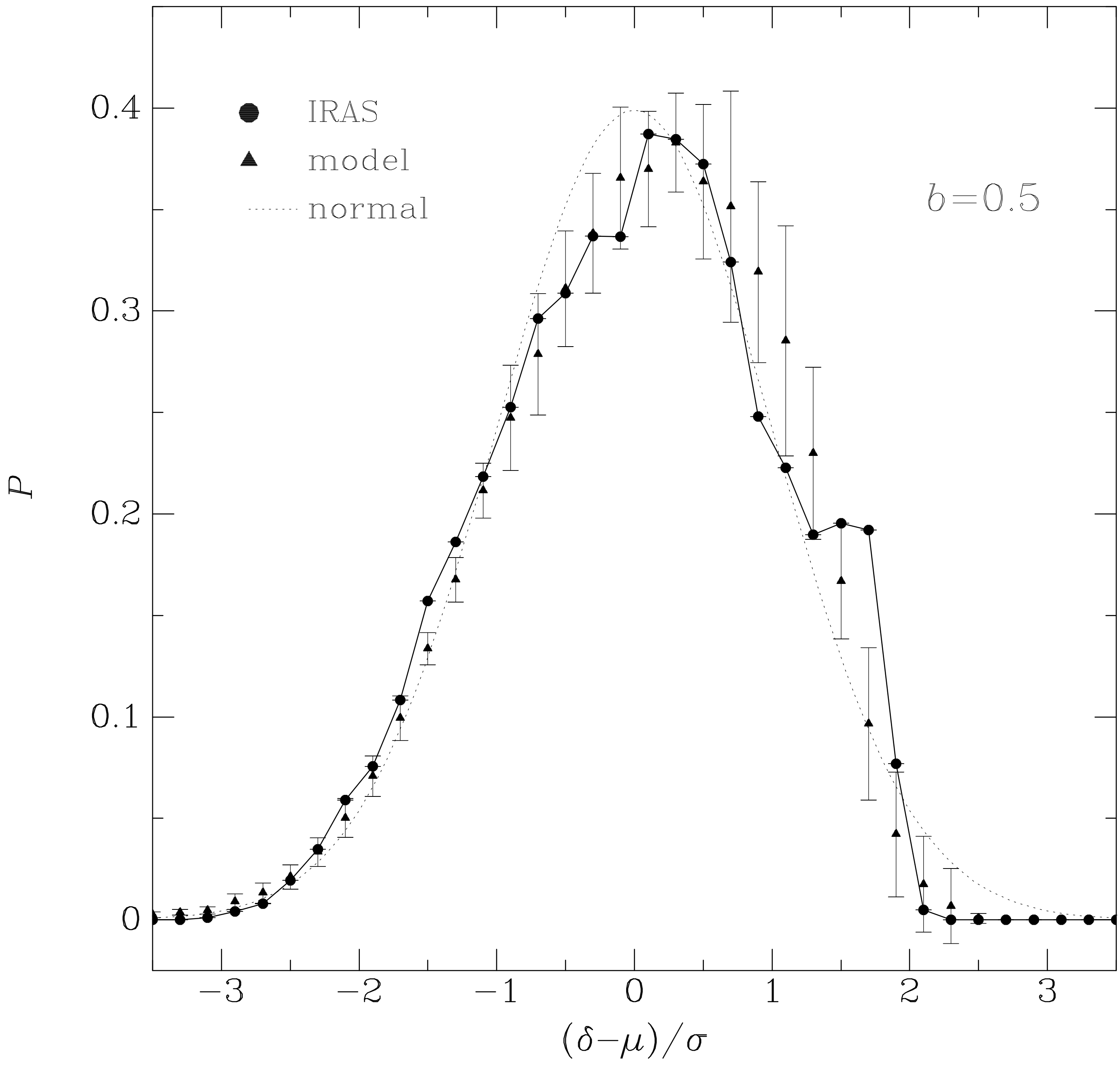

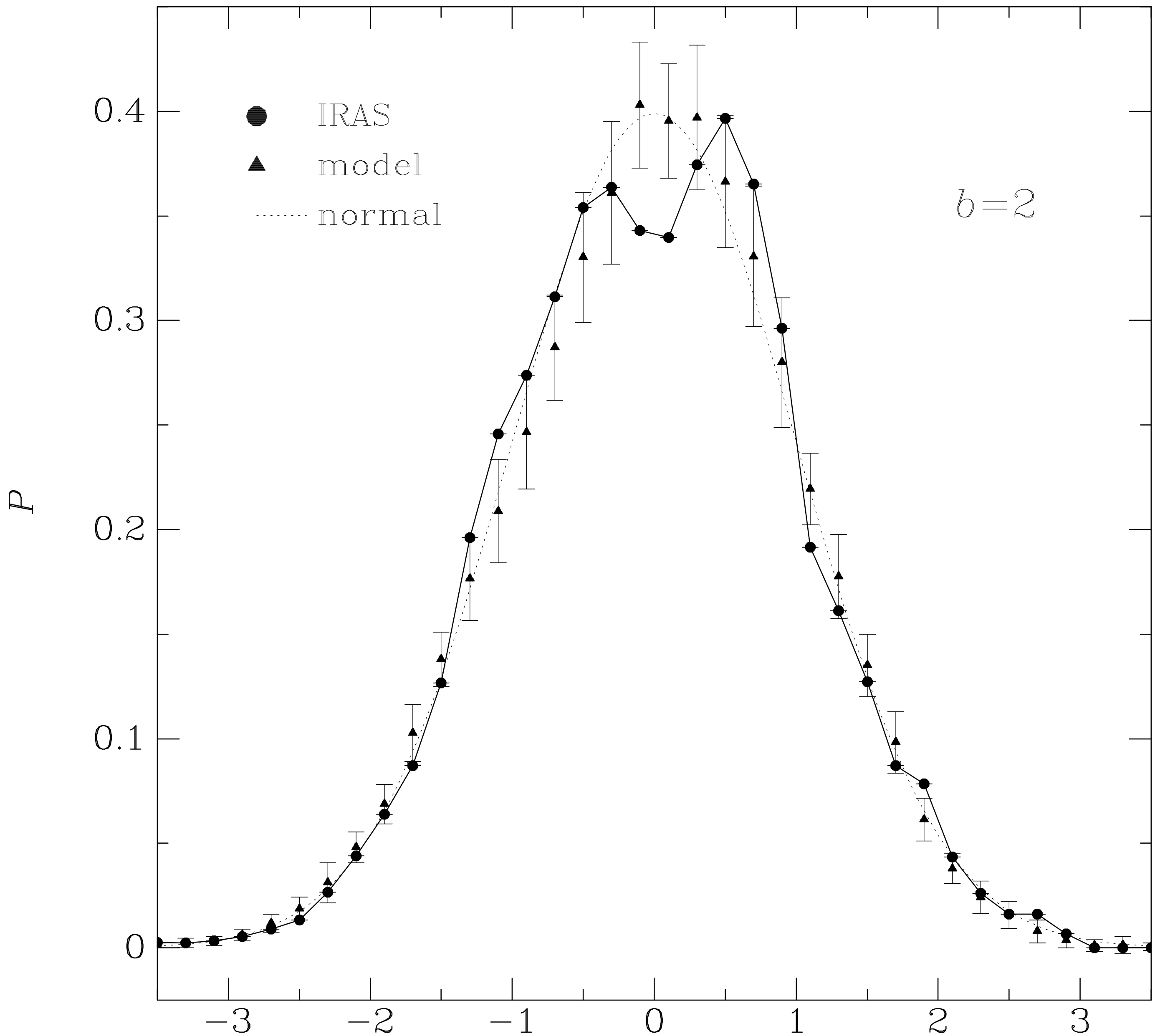

## FIGURE CAPTIONS

Fig. 1.— The PDF of the density field from the 1.2 Jy IRAS galaxy catalog (circles connected by a solid line) compared with a Gaussian (dotted line) and with the PDF of a model, computed from a series of 20 density fields of Gaussian CDM Monte-Carlo simulations (triangles). The error bars are the standard deviations over the simulations and measure both the effect of the limited volume sampled by IRAS and the effect of shot noise in the IRAS sampling. Note, however, that the data points are correlated due to smoothing.

Fig. 2.— The IPDF recovered from the 1.2 Jy IRAS catalog, compared to the IPDF recovered from the Gaussian CDM model, for different values of $b$. The notation is as in Fig 1.

# EVIDENCE FOR GAUSSIAN INITIAL FLUCTUATIONS FROM THE 1.2 JY IRAS SURVEY

Adi Nusser[1,2], Avishai Dekel[1], and Amos Yahil[3]

## ABSTRACT

We recover the one-point probability distribution function of the *initial* density fluctuations (IPDF) from the quasi-linear galaxy density field of the 1.2 Jy IRAS redshift survey smoothed by 10 $h^{-1}$Mpc. The recovery, using the laminar, Eulerian, Zel'dovich approximation of Nusser and Dekel, is independent of $\Omega$ and is relatively insensitive to linear galaxy biasing in the range $0.5 \le b \le 2$. Errors due to discrete sampling, the limited survey volume, and the method of recovering the IPDF are evaluated by comparing the IPDF determined from the IRAS data with Monte-Carlo IRAS-like catalogs "observed" from $N$-body simulations of CDM models with Gaussian initial conditions. Eight sensitive statistical tests used in this comparison find the IRAS IPDF to be consistent with Gaussian. We provide observational constraints on possible deviations from Gaussianity, which should be obeyed by any theoretical model.

*Subject headings:* cosmology — dark matter — galaxies: clustering — galaxies: formation — gravity — large-scale structure of the universe

[1]Racah Institute of Physics, The Hebrew University of Jerusalem, 91904, Israel.

[2]Center for Particle Astrophysics, University of California, Berkeley, CA 94720.

[3]Astronomy Program, State University of New York, Stony Brook, NY 11794-2100.

## 1. INTRODUCTION

The one-point initial probability distribution function (IPDF) is of particular interest among the statistics that characterize the primordial density fluctuation field. Whether or not it was Gaussian has far-reaching cosmological implications. First, the IPDF provides direct information about the origin of the fluctuations, being independent of the nature of the dark matter, unlike the power spectrum, whose evolution during the plasma era depends on the physics of the dark matter. Moreover, a known IPDF would place very interesting constraints on the value of the cosmological density parameter, $\Omega$, based on the observed large-scale peculiar velocity field, but independent of galaxy biasing (Nusser & Dekel 1993, henceforth ND; Bernardeau *et al.* 1993).

The "standard" model, based on inflation, predicts that the initial fluctuations were a Gaussian random field (to all orders of the joint IPDF's, cf., Bardeen *et al.* 1986), while competing models predict various kinds of non-Gaussian fluctuations. The latter include certain inflation models (cf., Kofman *et al.* 1991), as well as models where the perturbations were seeded by cosmic strings (see Bertschinger 1989 for a review), by textures (Turok 1991) or by non-gravitational explosions (see Ostriker 1988 for a review).

Even if the IPDF were Gaussian, the present probability distribution function (PDF) is expected to be non-Gaussian due to nonlinear effects. The tails of the distribution develop positive skewness because high peaks collapse to large densities while the density in voids cannot become negative. At moderate amplitudes the fluctuations develop skewness of an opposite sign because fluctuations above the mean tend to contract while expanding underdense regions occupy more comoving volume as time goes by.

The detailed quasi-linear evolution of the PDF in an initially Gaussian system has been studied using several approximations. For example, according to second-order perturbation theory, the third moment of the density distribution grows in proportion to the square of the variance (Peebles 1980; Bouchet *et al.* 1992; Juszkiewicz & Bouchet 1992; Coles & Frenk 1992; but see Lahav *et al.* 1993). Kofman *et al.* (1993), have found that the density PDF actually develops a log-normal shape, as originally proposed by Coles & Jones (1991). This is demonstrated in Fig. 1, in which we plot as triangles the mean of the PDF from 20 $N$-body simulations (§4), smoothed by a Gaussian radius of 10 $h^{-1}$Mpc, and as the long-dashed curve a log-normal distribution with the same standard deviation. The error bars are the dispersion per bin among the $N$-body simulations.

The present PDF of the density field derived from the 1.2 Jy IRAS galaxy catalog with the same smoothing (§2), is also shown in Fig. 1. The deviations of the present IRAS PDF from Gaussianity agree well with the $N$-body simulations. This is a necessary condition for Gaussian initial fluctuations, but it is important to realize that the present PDF has only limited discriminatory power against non-Gaussianity in the initial conditions. The PDF tends to develop a log-normal shape in a robust way even for certain non-Gaussian initial fluctuations (*e.g.* Weinberg & Cole 1992, Fig. 14).

A more effective strategy to determine the statistical nature of the initial fluctuations is to take advantage of the *full* density field available at the present epoch, trace it back in time to recover the initial fluctuation field (or certain statistics that characterize it), and use those to discriminate between theories in the linear regime. One way to accomplish this goal is by first recovering the full growing mode of the initial fluctuation field (e.g., Peebles 1989, 1990; Nusser & Dekel 1992; Giavalisco *et al.* 1993), and then computing the IPDF from it. Instead, we use below the simple short cut approximation developed by ND in which the IPDF is recovered directly from the present quasi-linear density field. We prefer not to limit the analysis to certain moments of the distribution, such as the skewness, because they might be dominated by the poorly determined tails of the distribution and might even diverge. Instead, we recover the full IPDF and only then use several different statistics to measure its characteristic features, which may or may not be emphasized by the standard moments.

The shape of the PDF is preserved in the linear regime, so a study of its evolution is necessarily nonlinear. On the other hand, we wish to avoid highly nonlinear systems in which multi-streaming and mixing obliterate the memory of initial conditions, thus rendering impossible the recovery of the IPDF. This restricts us to the quasi-linear regime, in which the approximation due to Zel'dovich (1970) is an excellent tool. We follow ND in using the Zel'dovich approximation to determine the IPDF from the present day velocity field, the latter being calculated from the density field using the phenomenological approximation of Nusser *et al.* (1991, hereafter NDBB).

The recovery of the IPDF from a given galaxy-density field requires, in principle, two assumptions: (1) a "biasing" relation between galaxy density and mass density and (2) a cosmological model, i.e., a value for $\Omega$. (The growth of perturbations is insensitive to the cosmological constant, e.g., Lahav *et al.* 1991.) However, in the Zel'dovich approximation which we use, the IPDF recovered from a present day *density* field does not depend on the assumed value of $\Omega$ (ND, see §3.6 below), unlike the IPDF determined from the present day velocity field, which is very sensitive to it (ND). Moreover, the dependence on biasing is also quite weak in practice. We are therefore able to recover the IPDF with great confidence, with little or no dependence on biasing and on the cosmological model.

The data are described in §2, the recovery method of the IPDF is detailed in §3, the use of $N$-body simulations for error analysis is discussed in §4, the recovered IPDF is shown and several statistics are computed in §5, and our conclusions are summarized in §6.

## 2. DATA

The database used here is a complete redshift survey of 5313 galaxies brighter than 1.2 Jy at 60$\mu$m detected by the Infrared Astronomical Satellite (*IRAS*).[4] Galaxy candidates were chosen from the *IRAS* Point Source Catalog, Version 2, (1988) using the selection criteria described in Strauss *et al.* (1990) and Fisher (1992). The data for the brighter half of the sample can be found in Strauss *et al.* (1992). At present, thirty objects (0.5% of the sample) remain unobserved. Sky coverage is complete for $|b| > 5^\circ$, with the exception of a small region of the sky which *IRAS* failed to survey and regions limited by confusion; our sample covers 87.6% of the sky.

The transformation of a redshift survey to a three-dimensional density map involves a number of ingredients. Full details are given by Yahil *et al.* (1991). The important steps are an accurate determination of the selection function, i.e., the probability that a galaxy at a given distance be included in the sample, correction for the 12% of the sky not covered by the survey, and a self-consistent transformation from the observed redshifts to distances, using peculiar velocities deduced from the galaxy distribution itself. This transformation from *redshift* space to real space, in which redshifts are corrected for peculiar velocities, is calculated in linear theory, assuming $\Omega^{0.6}/b = 1$. Method 2 of Yahil *et al.* (1991) is used in the present work.

Our treatment is not completely self-consistent because the transformation from redshifts to real distances for the *IRAS* galaxies was determined using *linear* theory (Yahil *et al.* 1991), while the velocity field used for the purpose of the recovery of the *IPDF* in the present paper is calculated from the *IRAS* density field using *nonlinear* theory. However, the smoothed *IRAS* density field is not very sensitive to this detail.

The density field of the 1.2 Jy *IRAS* galaxies described above was first computed using a cloud-in-cell procedure on a cubic grid measuring 240 $h^{-1}$Mpc on the side, centered on the Local Group, with a spacing of 5 $h^{-1}$Mpc. It was normalized to an average density of unity in a sphere of radius 70 $h^{-1}$Mpc centered on the Local Group. The density was then smoothed on the full grid by a Gaussian with radius $R_s = 10$ $h^{-1}$Mpc using FFT. In the recovery of the *IPDF* (§3.1–§3.4), however, only a sub-grid of 160 $h^{-1}$Mpc centered on the Local Group was used. In order to reduce errors due to sparse sampling, the final sampling of the *IPDF* (§3.6) was then further restricted to an inner sphere of radius 70 $h^{-1}$Mpc centered on the Local Group.

---

[4]The Infrared Astronomical Satellite was developed and operated by the U.S. National Aeronautics and Space Administration (NASA), the Netherlands Agency for Aerospace Programs (NIVR), and the U.K. Science and Engineering Research Council (SERC).

## 3. RECOVERY METHOD

The ND recovery method begins with a smoothed density field of galaxies on an Eulerian grid at the present epoch and determines the IPDF in the following steps: (a) convert the galaxy density field observed at the present epoch to a mass density fluctuation field, $\delta(\boldsymbol{x})$, using an assumed "biasing" relation, (b) solve for the associated peculiar velocity field, $\boldsymbol{v}(\boldsymbol{x})$ using a nonlinear scheme, (c) compute the deformation tensors with respect to the Eulerian coordinates at the present epoch at the grid positions, (d) use the Zel'dovich approximation to convert them to deformation tensors with respect to the corresponding Lagrangian coordinates, (e) obtain the initial linear densities at the above Lagrangian coordinates from the Lagrangian deformation tensor, and, (f) use the initial linear densities at the Lagrangian coordinates to compute the volume—i.e., mass—weighted distribution of density at that epoch.

### 3.1. From Galaxy to Mass density

If galaxy formation is biased, galaxies need not be faithful tracers of the mass (see Dekel & Rees 1987 for a review). There is, however, growing evidence that the galaxy and mass densities are strongly correlated (Dekel *et al.* 1993). Both fields feature the same main structures—the Great Attractor, the Perseus-Pisces supercluster and the large void in between—and they are consistent with being noisy versions of each other.

In this paper we assume a deterministic power-law relation between the galaxy and mass densities,

$$1 + \delta_I = (1 + \delta)^b \quad , \tag{1}$$

where $\delta_I$ is the smoothed density fluctuation field of IRAS galaxies and $\delta$ is the mass-density fluctuation field. In the limit $\delta \ll 1$, this model reduces to linear biasing, $\delta_I = b\delta$, which is the simplest realization of a linear statistical biasing relation between the variances of the fields, as is roughly predicted for linear density peaks in a Gaussian field (Kaiser 1984; Bardeen *et al.* 1986). The power law relation attempts to give a better description of biasing, particularly at negative $\delta$'s, where linear biasing can lead to $\delta < -1$. The nonlinear relation between $\delta_I$ and $\delta$ makes the mean value, $\langle\delta\rangle$, deviate slightly from zero (the IRAS density is normalized so $\langle\delta_I\rangle = 0$), but this small shift would not affect the IPDF, which is computed relative to the mean in the given volume.

The correlation between the density fields of galaxies and mass has been found by Dekel *et al.* (1993) to be consistent with the above biasing scheme. Motivated by this result, we confine the following analysis to $0.5 \leq b \leq 2$, their 95% confidence limits for $\Omega < 1.5$. Our analysis is independent of the value of $\Omega$ (§3.6); in practice we use $\Omega = 1$.

### 3.2. From Density to Velocity

In the linear regime the velocity divergence is simply $\boldsymbol{\nabla}\cdot\boldsymbol{v} = -\dot{a}f(\Omega)\delta$ where $f(\Omega) = d\log D/d\log a \approx \Omega^{0.6}$ is the logarithmic growth rate of the density perturbations with respect to the expansion scale, and the gradient is with respect to the comoving coordinates. NDBB found empirically that in the quasi-linear regime this divergence is well approximated by

$$\boldsymbol{\nabla}\cdot\boldsymbol{v} = \frac{-\dot{a}f(\Omega)\delta}{1+0.18\delta} \quad . \tag{2}$$

Eq. (2) is a Poisson equation in which the density has been replaced by a simple function of the density, and since the smoothed velocity field is irrotational for both linear and quasi-linear perturbations (e.g, Dekel, Bertschinger, & Faber 1990), it can be computed with the usual grid-based FFT techniques. In order to reduce the effect of the periodic boundary conditions, we "zero pad" by embedding our 160 $h^{-1}$Mpc grid in one that measures 320 $h^{-1}$Mpc on the side, with $\delta = 0$ in the added volume. We do not use the *IRAS* data which we do have in that region, because our smoothing length of 10 $h^{-1}$Mpc becomes smaller there than the mean distance between observed galaxies, and the nonlinear correction in Eq. (2) can lead to large errors.

The accuracy of the velocity field derived using this approximation was tested by NDBB and more recently by Mancinelli, Yahil, Ganon & Dekel (1993) in $N$-body simulations. With Gaussian smoothing of 10 $h^{-1}$Mpc, the approximation was found to have *rms* errors less than 50 km $s^{-1}$, compared with the peculiar velocities themselves, which can exceed 1000 km $s^{-1}$.

### 3.3. Eulerian Deformation Tensor

The Zel'dovich approximation assumes that the comoving Eulerian position $\boldsymbol{x}$ of a particle with initial Lagrangian position $\boldsymbol{q}$ can be written as

$$\boldsymbol{x}(\boldsymbol{q},t) = \boldsymbol{q} + D(t)\boldsymbol{\psi}(\boldsymbol{q}) \quad , \tag{3}$$

where the displacement is separable into a product of the growth function $D(t)$ and a time-independent spatial perturbation function, $\boldsymbol{\psi}$. It follows that the velocity is given by

$$\boldsymbol{v} = a\dot{\mathrm{x}} = a\dot{D}\boldsymbol{\psi} \quad . \tag{4}$$

We assume further that the flow is laminar, and that there exists a one-to-one correspondence between the Eulerian and Lagrangian positions. The latter can therefore be expressed in terms of the Eulerian position and the velocity as

$$\boldsymbol{q}(\boldsymbol{x},t) = \boldsymbol{x} - D\boldsymbol{\psi}(\boldsymbol{x},t) \quad . \tag{5}$$

The Eulerian deformation tensor, $\partial\psi_i/\partial x_j$, is readily computed on the grid from the velocities determined in §3.2, using Eq. (4) to relate $\boldsymbol{v}$ and $\boldsymbol{\psi}$. The Lagrangian deformation tensor, $\partial v_i/\partial q_j$, can be obtained from the Eulerian one, using the chain rule, to give

$$\frac{\partial\psi_i}{\partial x_j} = \left[\delta_{kj} - D\frac{\partial\psi_k}{\partial x_j}\right]\frac{\partial\psi_i}{\partial q_k} \quad , \tag{6}$$

where the expression in the square bracket is the Jacobian matrix obtained from Eq. (5).

### 3.4. Initial density

Eq. (6) can be inverted to yield the full Lagrangian deformation tensor. In fact, the initial linear density depends only on its trace

$$\delta_{in} = -D_{in}\frac{\partial\psi_i}{\partial q_i} \quad . \tag{7}$$

For irrotational flow the deformation tensor is symmetric, and its trace therefore equals the sum of its eigenvalues. Moreover, Eq. (6) clearly shows that the Eulerian deformation tensor and the Jacobian transformation matrix can be simultaneously diagonalized. Hence, each eigenvalue of the Lagrangian deformation tensor is given by

$$\lambda_i = \frac{\mu_i}{1 - D_0\mu_i} \quad , \tag{8}$$

where $\mu_i$ is the equivalent eigenvalue of the Eulerian deformation tensor at the present epoch. The initial density can therefore be written as

$$\delta_{in} = -\frac{D_{in}}{D_0}\sum_i \frac{D_0\mu_i}{1 - D_0\mu_i} \quad . \tag{9}$$

### 3.5. From Initial Density to IPDF

In the last step, we wish to compute the IPDF of $\delta_{in}$, $P(\delta_{in})$, i.e., the distribution of $\delta_{in}$ over equal volume elements, which in the linear limit means equal mass elements. But the points at which we sample the present density and computed the initial density are homogeneously distributed at the present epoch, and not at the initial time. In order to correct for this inhomogeneous sampling, we assign each point $\boldsymbol{q}_n$ a weight $1 + \delta(\boldsymbol{x}_n)$, which, by mass conservation, is the relative mass which should be assigned to the Lagrangian position $\boldsymbol{q}_n$. In order to avoid errors due to the sparsity of IRAS sources at large distances, we restrict the sample for which we compute the IPDF to those points whose present positions, $\boldsymbol{x}_n$, are within a distance of 70 $h^{-1}$Mpc from the Local Group.

Note that there is no guarantee that the mean initial density we compute, $\langle\delta_{in}\rangle$, averages to zero. This is because we imposed the condition $\langle\delta\rangle = 0$ (more precisely $\langle\delta_I\rangle = 0$) on a finite volume at the present epoch, while, in fact, that volume may be a little overdense or underdense. We could iteratively correct our normalization until $\langle\delta_{in}\rangle = 0$. Instead we simply shift the $\delta_{in}$'s we obtained until their mean equals zero. Dekel (1981, the appendix) discusses the justification for this procedure.

### 3.6. Ω Dependence

Eq. (9) shows that, apart from overall scale factors, the dependence on $\Omega$ can enter the initial density $\delta_{in}$ only via the term $D_0\mu_i$. But Eq. (2) depends on $\Omega$ only through the scale factor $\dot{a}f$, and Eq. (4) introduces another factor of $1/a\dot{D}$. Hence, the quantities $D_0\mu_i$ are given by the product of a function of $\delta$ and $D_0\dot{a}f/a\dot{D}_0 \equiv 1$. The weights used to convert the $\delta_{in}$'s to the final IPDF (§3.5) are functions of $\delta$, so are also independent of $\Omega$. The dependence on $\Omega$ enters only in the transformation of the original IRAS density from redshift to configuration space, and this dependence is quite weak. In our approximation, therefore, the IPDF is almost independent of $\Omega$. There is a dependency on biasing, introduced in the transformation from $\delta_I$ to $\delta$, but this dependency is also weak (§5).

## 4. ERROR ANALYSIS

The derived IPDF can be in error compared with the universal IPDF for three main reasons: (a) the fact that the local volume sampled by the 1.2 Jy IRAS survey may still not be big enough to provide a "fair" sample, (b) shot noise due to the discrete sampling, which gets worse at larger distances from the Local Group, and (c) errors in the recovery procedure itself. We estimate all these errors simultaneously by computing the IPDF from many Monte-Carlo fake catalogs, which were "observed" *à la* IRAS from many random simulations of a cosmological model.

The $N$-body simulations were run using a particle-mesh code (Bertschinger & Gelb 1991) with $64^3$ grid cells and particles in a periodic cubic box of comoving size 160 $h^{-1}$Mpc. The initial conditions for each simulation were a random Gaussian realization of the "standard" CDM spectrum, with $\Omega = 1$ and $h = 0.5$ (Davis *et al.* 1985). The simulations were halted at a time step when, based on linear growth, the *rms* density fluctuation in top-hat spheres of radius 8 $h^{-1}$Mpc equaled a desired value, $\sigma_8$. Since the corresponding *rms* fluctuation observed for the galaxy distribution is roughly $b\sigma_8 = 1$, the value of $\sigma_8$ at the final step of the simulation equals the inverse of the biasing parameter, $b$. We ran 20 simulations for each of the three values $\sigma_8^{-1} = b = 0.5, 1, 2$.

We need to construct fake IRAS-like samples from each $N$-body simulation, in order to compare with the IRAS reconstruction. We create these samples using the following steps: (a) create a smoothed mass density field on a grid by cloud-in-cell, followed by Gaussian smoothing with a radius which equals the radius used in the IRAS reconstruction (10 $h^{-1}$Mpc), (b) bias the mass density field according to Eq. (1), (c) deconvolve the biased density field to obtain an unsmoothed galaxy density field,[5] (d) sample the "observed" galaxy number density field at each grid point by seeking a random Poisson deviate whose expectation value equals the galaxy density obtained in (c), multiplied by $n_1\phi(r)$, the expected number of galaxies which IRAS would observe at distance $r$ in a homogeneous universe (Yahil *et al.* 1991), (e) divide the galaxy number density at each grid point by $n_1\phi(r)$ to obtain a new "IRAS-noisy" distance-independent estimate of the real galaxy density. This provides a fake IRAS catalog, equivalent to the one obtained from the real data by the cloud-in-cell procedure. The first smoothing and deconvolution, steps (a)–(c), are needed only for biased models, because the biasing has to be applied to the smoothed density. On the other hand, the sampling using the IRAS selection function, steps (d)–(e), has to be carried out with the unsmoothed distribution, because the IRAS simulation so obtained will then be smoothed in the IPDF reconstruction, and we have to avoid double smoothing.

In Fig. 1, we show the present PDF of the initially-Gaussian model, with $b = 1$. The PDF is computed for each of the 20 Monte Carlo galaxy density fields. The mean of the 20 PDF values in each bin is shown as a triangle, and the standard deviation represents the error.

Each of the Monte Carlo galaxy density fields is then used as input for the recovery of the IPDF. The mean and standard deviation over the Monte Carlo simulations represent the model IPDF and the uncertainties about it. Fig. 2 shows the IPDF recovered from the model, and the associated errors, for the three different values of the biasing factor, $b = 0.5, 1, 2$, which are the "true" values used in the simulations themselves.

## 5. RESULTS

The IRAS galaxy density field is fed into the recovery procedure, which yields the IPDF shown as solid curves in Fig. 2 for the three assumed values of $b$. We can see that the IPDF is quite insensitive to the assumed value of $b$ in the relevant range $[0.5, 2]$. This insensitivity to the assumed $b$, allows us, without loss of generality, to compare the IRAS IPDF only to model IPDF's in which the recovery assumed the "true" value of $b$. We clearly see in Fig. 2 that the IRAS IPDF is consistent with a Gaussian, for all the three values of $b$ tested.

[5]The deconvolution is unstable if the broadening function in $k$-space is allowed to go to zero at high $k$. We modify the $k$-space Gaussian by multiplying it by 0.99 and adding a 0.01 white background.

We next wish to quantify the agreement, or discrepancy, between the data and the Gaussian model. Quantitative limits on the possible deviation from Gaussianity, by several different statistics, would be particularly useful for testing non-Gaussian theoretical models. The counts in the different bins that construct the IPDF are correlated in a complex fashion because the fields were smoothed over a large scale at the present epoch before the reconstruction was begun. In practice, such complex correlations can only be analyzed by using statistics that measure certain properties of the IPDF, and evaluate the distributions of these statistics using Monte Carlo simulations. We determine the standard deviation of each statistic over the Monte-Carlo catalogs, and assume a Gaussian distribution for this statistic when we express the confidence level of rejection in terms of numbers of sigma.

The results of several statistics are summarized in Table 1, which lists for each value of $b$ and for each statistic the mean value and the standard deviation in the Gaussian model, the value measured from the IRAS data, and its deviation from the value obtained for the Gaussian model in units of the standard deviation.

*[Table 1]*

First listed are $\mu$ and $\sigma$, the mean and standard deviation of $\delta_{in}$. Then comes $W$, a standard statistic which has been optimized to measure deviations from Gaussianity when the distribution is sampled by a small number of independent points. Given the initial densities at grid points, $\delta_i$, $i = 1, n$, the statistic is defined by

$$W = \frac{1}{(n-1)\sigma^2} \left( \sum_{i=1}^{[n/2]} a_i (\delta_{n+1-i} - \delta_i) \right)^2 \quad , \tag{10}$$

where $[n/2]$ indicates the integer part of $n/2$, and the coefficients $a_i$ are chosen to maximize the discriminatory power against Gaussianity (Shapiro & Wilk 1965). The coefficients are computed according to the prescription described by these authors. The initial density at each grid point was weighted by the present density there.

There follows a $\chi^2$-type statistic, which measures a vertical deviation between the data IPDF and the model IPDF:

$$\chi^2 = \sum_i \frac{[P_i(data) - P_i(model)]^2}{\sigma_i^2} \quad , \tag{11}$$

where $i$ runs through all the bins and $\sigma_i$ is the total error of $P_i$ in that bin. This is, of course, not a proper $\chi^2$ because the data points are correlated.

The rest are higher moments of the distribution. Let $x \equiv (\delta_{in} - \mu)/\sigma_{\delta_{in}}$. By definition $\langle x \rangle = 0$ and $\langle x^2 \rangle = 1$. The skewness and kurtosis are commonly defined by

$$S \equiv \langle x^3 \rangle, \quad K \equiv \langle x^4 \rangle - 3 \ , \tag{12}$$

such that they both vanish for a Gaussian distribution. We see from Table 1 that these moments carry large random errors, on the order of a few tenths. This is because they are dominated by the poorly determined tails of the distribution. For measuring similar properties of the IPDF but with less weight assigned to the tails, ND suggested replacing the skewness and kurtosis by

$$S_r \equiv \langle x|x|\rangle, \quad K_r \equiv \langle |x|\rangle - \sqrt{\frac{2}{\pi}} \quad . \tag{13}$$

The former, ranging from $-1$ to $+1$, measures mirror asymmetry about zero. The latter, ranging from $-\sqrt{2/\pi}$ to $1-\sqrt{2/\pi}$, measures symmetric deviations from a Gaussian; it is negative when the PDF is dominated by excessive tails or by a shallow center. We see in Table 1 that the errors of these moments are much better behaved; their random errors are only on the order of a few percents.

As expected on the basis of a visual inspection of Fig. 2, the Gaussian model cannot be ruled out by the data using any of the statistics. The most discriminatory among the statistics seems to be $W$, by which the model with $b=1$ is "rejected" at the 1.85-sigma level. In all other cases the IRAS datum lies well within the very likely region of the distribution of the statistic. The IPDF is thus fully consistent with a normal curve. This conclusion is insensitive to the value of $b$, at least in the range $[0.5, 2]$.

In order to quote n-sigma limits on the allowed deviation from Gaussianity based on any statistic $s$, one can compute from Table 1 the quantity $(s-\langle s\rangle) \pm n\sigma_s$, where $\langle s\rangle$ and $\sigma_s$ are the model Monte-Carlo mean and standard deviation of $s$. For example, For a $b=1$ model we find the 3-sigma constraints, $0.87 < W < 1.03$, $\chi^2 < 1.2$, $-0.18 < S_r < 0.11$, $-0.205 < K_r < -0.097$, $-0.65 < S < 0.36$, and $-0.82 < K < 0.62$.

## 6. DISCUSSION AND CONCLUSIONS

Based on the 1.2 Jy IRAS redshift catalog, and for biasing-parameter values in the range of interest, $0.5 \leq b \leq 2$, the density IPDF is fully consistent with a Gaussian. Any possible deviations from a Gaussian are confined by several sensitive statistics to a small permitted range; these statistics could also be useful when trying to evaluate the likelihood of non-Gaussian models. The method used here, based on a recovery of the IPDF, is more sensitive than tests using the present PDF.

One limitation of the current results is the model dependence of the error analysis. in the Monte-Carlo analysis, we had to use a Gaussian model with a specific power spectrum. Although the power spectrum is expected to have a negligible effect on the IPDF itself, it should affect, for example, the estimated errors due to the limited volume sampled. The CDM spectrum used here is an appropriate choice because it is consistent with the observed power spectrum over the scales used in this investigation (Vogeley *et al.* 1992; Fisher *et al.* 1993; Baugh & Efstathiou 1993; Kolatt & Dekel 1994). It would nevertheless be worthwhile to test the method and estimate the errors using other models with different spectra.

Another limitation is that the constraints on possible deviations from a Gaussian *IPDF* were derived based on Monte-Carlo simulations of a Gaussian model. The appropriate procedure for testing a specific non-Gaussian model is to estimate the distribution of each statistic by Monte-Carlo simulations perturbed from that specific model. The constraints provided by Table 1 are therefore reliable only for models that do not deviate drastically from Gaussian. Any testing of a strongly non-Gaussian model would require a repetition of the full Monte-Carlo analysis using the ND recovery technique.

The Gaussian *IPDF* found here from *IRAS* adds a crucial ingredient to the framework discussed in ND. Quasi-Linear gravity is used to relate two observables: (1) a smoothed velocity field and (2) a smoothed galaxy-density field, to three fundamental theoretical ingredients: (a) the cosmological model via $\Omega$, (b) the relation between galaxy and mass density via the biasing parameter $b$, and (c) the initial fluctuations via their *IPDF*. Given a galaxy-density field, one can recover the *IPDF* independently of $\Omega$ and with only weak dependence on $b$. Given a velocity field, one can assume a value for $\Omega$ and obtain the *IPDF*, independently of biasing. Given both fields, one can simultaneously constrain both $\Omega$ and the *IPDF*.

The contribution of the current analysis is to establish the Gaussian nature of the initial fluctuations, essentially independently of the assumed values of $\Omega$ and $b$. This, in turn, argues that $\Omega$ is about unity or larger, because ND showed that the *POTENT* velocity field is consistent with a Gaussian *IPDF* *if and only if* $\Omega$ has such a large value. For example, values of $\Omega \leq 0.3$ were ruled out at the 4-6 sigma level by several different statistics. Similar, though somewhat weaker conclusions, were reached by Bernardeau *et al.* (1993) based on the skewness of $\boldsymbol{\nabla} \cdot \boldsymbol{v}$ in the *POTENT* reconstruction and the assumption of Gaussian initial fluctuations.

Another two of the theoretical ingredients, $\Omega$ and $b$, can be determined from a direct quasi-linear comparison of the *POTENT*-mass and *IRAS*-galaxy density fields. The strongest result of Dekel *et al.* (1993) is $\Omega^{0.6}/b = 1.28^{+0.75}_{-0.59}$ at the 95% confidence limit. Separate nonlinear constraints on $\Omega$ and $b$, which are somewhat weaker, are $\Omega > 0.46$ for $b > 0.5$ and $b = 0.7^{+0.6}_{-0.2}$ for $\Omega = 1$, again at the 95% confidence level. These constraints, as well as the ones from the *POTENT* reconstruction of the *IPDF*, are expected to tighten shortly with forthcoming new velocity data.

Once the analysis is sufficiently nonlinear, we expect only one simultaneous solution for all three theoretical ingredients. The solution we have found so far from the above-mentioned investigations combined, i.e., $\Omega \gtrsim 1$, $b_I \lesssim 1$, and Gaussian *IPDF*, are all consistent. The constraints therefore seem to be closing in on a "standard" model of an Einstein–de Sitter flat universe with Gaussian initial fluctuations and *IRAS* galaxies roughly tracing mass on scales $\sim 10\ \mathrm{h^{-1}Mpc}$. Low values of $\Omega$, with or without a cosmological constant, or non-Gaussian initial fluctuations, cannot be easily invoked when trying to explain the remaining puzzles of the formation of large-scale structure in the universe.

We thank E. Bertschinger for his $N$-body code, and the Institut d'Astrophysique in Paris and the Observatoire de Paris in Meudon for their hospitality while most of this work was done. This research has been supported by US-Israel Binational Science Foundation grants 89-00194 and 92-00355 and NASA grant NAG 51228.

Table 1. Statistics of the *IPDF* from *IRAS*

| | **b = 0.5** | | | **b = 1** | | | **b = 2** | | |
|---|---|---|---|---|---|---|---|---|---|
| Statistic | Model | *IRAS* | Deviation | Model | *IRAS* | Deviation | Model | *IRAS* | Deviation |
| $\mu$ | $-0.014 \pm 0.094$ | $-0.016$ | $-0.27$ | $-0.070 \pm 0.031$ | $-0.045$ | $+0.81$ | $-0.002 \pm 0.006$ | $-0.008$ | $-0.94$ |
| $\sigma$ | $+0.321 \pm 0.024$ | $+0.306$ | $-0.65$ | $+0.650 \pm 0.070$ | $+0.560$ | $-1.29$ | $+0.177 \pm 0.013$ | $+0.170$ | $-0.49$ |
| $W$ | $+0.948 \pm 0.026$ | $+0.900$ | $-1.85$ | $+0.886 \pm 0.095$ | $+0.867$ | $-0.20$ | $+0.972 \pm 0.017$ | $+0.953$ | $-1.00$ |
| $\chi^2$ | $+0.535 \pm 0.211$ | $+0.445$ | $-0.42$ | $+0.610 \pm 0.280$ | $+0.810$ | $+0.69$ | $+0.621 \pm 0.445$ | $+0.530$ | $-0.20$ |
| $S_r$ | $-0.036 \pm 0.048$ | $-0.044$ | $-0.15$ | $-0.110 \pm 0.060$ | $-0.050$ | $+1.00$ | $-0.032 \pm 0.060$ | $-0.009$ | $+0.36$ |
| $K_r$ | $-0.151 \pm 0.018$ | $-0.132$ | $+1.00$ | $-0.143 \pm 0.022$ | $-0.124$ | $+0.87$ | $-0.159 \pm 0.021$ | $-0.143$ | $+0.75$ |
| $S$ | $-0.145 \pm 0.168$ | $-0.156$ | $-0.06$ | $-0.478 \pm 0.268$ | $-0.185$ | $+1.00$ | $-0.103 \pm 0.225$ | $-0.033$ | $+0.31$ |
| $K$ | $-0.100 \pm 0.240$ | $-0.360$ | $-1.01$ | $+0.590 \pm 1.420$ | $-0.560$ | $-0.81$ | $-0.018 \pm 0.370$ | $-0.091$ | $-0.19$ |